\documentstyle{article}

\begin{document}

\title{Bosonic Realization of Boundary Operators in\\
$SU(2)$-invariant Thirring Model}

\author{
Liu Chao$^{\ddag,\dagger}$ \hskip 0.5truecm Boyu Hou$^{\dagger}$
\hskip 0.5truecm Kangjie Shi$^{\dagger}$ \\
Yanshen Wang$^{\ddag,\dagger}$ \hskip 0.5truecm
Wenli Yang$^{\ddag,\dagger}$ \\
\bigskip\\
$^{\dagger}$ Institute of Modern Physics, Northwest
University, Xian 710069, China
\thanks{Mailing address}\\
$^{\ddag}$ CCAST (World Laboratory), P O Box 8730, Beijing 100080, China
}

\maketitle

\begin{abstract}
Boundary operators and boundary states in $SU(2)$-invariant Thirring model
are considered from the point of view of bosonization and oscillator
realizations of bulk and boundary Zamolodchikov-Faddeev algebras.
\end{abstract}

\section{Introduction}

The integrability of quantum theory implies that one can obtain the
exact correlation function or form factors for arbitrary many local operators
in the theory. The recent investigations on such theories have lead to
important progresses. In exact solvable lattice models \cite{1}-\cite{2d},
the free boson realization of $q$-deformed affine algebras
\cite{4} has provided a bosonization
of type $I$ and type $II$ vertex operators \cite{5}-\cite{4c} which satisfy
Zamolodchikov-Faddeev (ZF) algebra, and hence the correlation functions
of six-vertex model or
$XXZ$-chain are successfully calculated. In quantum field theories, {\it e.g.}
Sine-Gordon and $SU(2)$-invariant Thirring models, Lukuyanov \cite{6}
constructed two kinds of generators of ZF algebra--using free bosons and
$q$-deformed oscillators \cite{a,aa}--which behave quite similar to the
type $I$ and type $II$ operators mentioned above, and also
calculated the form factors \cite{7,7a} in these theories.
Another important aspect of recent progresses in integrable models is the
work of Ghoshal and Zamolodchikov \cite{8}, which proposed the
theories of integrable boundaries and boundary operators. Jimbo
{\it et al} \cite{9,9a} obtained the
$q$-boson realization for the boundary operators of $XXZ$-model and the
corresponding multipoint correlation functions, and it is the present
work which is intended to the study of boundary operators and boundary
states in $SU(2)$-invariant Thirring model and their boson realization.

\section{$SU(2)$-invariant Thirring model in the bulk: an overview}

Let us first review Lukuyanov's boson realization of
ZF algebra and the corresponding $S$-matrix in the case of $SU(2)$-invariant
Thirring model \cite{6}.

As mentioned in the introduction, in integrable
quantum  field theories, there are two kinds of operators, {\it i.e.} local
and asymptotic operators, among each there is a subset satisfying ZF algebra.

The local generators $Z'_a(\beta)$ (here and after we shall adopt the
convention that operators with a prime denote local operators and those
without a prime denote asymptotic operators) satisfy the ZF algebra

\begin{equation}
Z'_a(\beta_1)Z'_b(\beta_2)=S'^{cd}_{ab}(\beta_1 - \beta_2)
Z'_d(\beta_2)Z'_c(\beta_1) \label{1},
\end{equation}

\noindent where the $S$-matrix ${S'}^{cd}_{ab}(\beta)$ is determined by
the well-known Yang-Baxter equation, unitarity and crossing symmetry,
and summation over repeated indices is
understood. Explicitly, the $S$-matrix reads

\begin{eqnarray*}
\displaystyle
S'^{cd}_{ab}(\beta)&=&S'(\beta) \left(
\begin{array}{cccc}
1  &                           &                           & \cr
   &\frac{-\beta}{i\pi + \beta}&\frac{i\pi}{i\pi + \beta}  & \cr
   &\frac{i\pi}{i\pi + \beta}  &\frac{-\beta}{i\pi + \beta}& \cr
   &                           &                           &1
\end{array}
\right),\\
\displaystyle
S'(\beta)&=&\frac{\Gamma\left(\frac{i\pi-\beta}{2i\pi}\right)
\Gamma\left(\frac{2i\pi+\beta}{2i\pi}\right)}{\Gamma\left(
\frac{i\pi+\beta}{2i\pi}\right)\Gamma\left(\frac{2i\pi-\beta}{2i\pi}\right)}.
\end{eqnarray*}

The ZF algebra (\ref{1}) can be realized through the bosonic field
$\phi'(\beta)$ with the commutation relation

\begin{eqnarray}
\left[\phi'(\beta_1),\;\phi'(\beta_2)\right]
= {\rm ln} S'(\beta_2-\beta_1),\label{phiphi}
\end{eqnarray}

\noindent as follows,

\begin{eqnarray*}
Z'_+(\beta)&=&V'(\beta)\equiv :{\rm e}^{i\; \phi'(\beta)}:,\\
Z'_-(\beta)&=&i\;(\chi'V'(\beta)+V'(\beta)\chi'),
\end{eqnarray*}

\noindent where $\beta$ is the rapidity of the boson $\phi'$, {\it i.e.}
the momentum $p'$ of $\phi'$ is related to $\beta$ via $p'=(p_0,\;p_1)
=m'({\rm ch}\beta,\;{\rm sh} \beta)$, $\chi'$ is defined through

\begin{eqnarray*}
\langle u | \chi' | v \rangle &\equiv& \eta' \langle u | \int_{C'}
\frac{{\rm d}\;\gamma}{2 \pi} \bar{V}'(\gamma) | v \rangle,\\
\bar{V}'(\gamma) &\equiv& :{\rm e}^{-i\;\bar{\phi}'(\gamma)}:,\\
\bar{\phi}'(\gamma) &=& \phi'(\gamma+\frac{i\;\pi}{2})
+ \phi'(\gamma-\frac{i\;\pi}{2}),
\end{eqnarray*}

\noindent in which $\langle u |$ and $| v \rangle$ are some states in
the (dual) Fock space of the boson $\phi'$, $\eta'$ is an irrelavent constant,
and the integration
contour $C'$ is taken such that it encloses only the poles originated
from the action of $\bar{V}'$ on the right-handed state $| v \rangle$
clockwise.

Notice that, instead of considering the actions of local operators onto
the Hilbert space ${\cal H}$, we now consider their actions on the
Fock space ${\cal F}$ which is a proper subspace of ${\cal H}$.
To specify the Fock space ${\cal F}$ more clearly, one needs the unique
$SU(2)$ invariant vacuum state $|0 \rangle$ under which the two-point
function of the boson $\phi'(\beta)$ reads

\begin{eqnarray}
\langle 0 | \phi'(\beta_1)\phi'(\beta_2) | 0 \rangle
= - {\rm ln} g'(\beta_2-\beta_1).\label{2}
\end{eqnarray}

\noindent The consistence of (\ref{phiphi}) with (\ref{2})
implies that $g'(\beta)$
is the Riemann-Hilbert factor of $S'(\beta)$ which is analytic in the
lower half plane ($\Im m \beta < 0$),

\begin{eqnarray*}
S'(\beta)= \frac{g'(-\beta)}{g'(\beta)},\hskip 1truecm
g'(\beta)= k^{1/2}
\frac{\Gamma \left(\frac{2i\pi-\beta}{2i\pi}\right)}{\Gamma
\left(\frac{i\pi-\beta}{2i\pi}\right)}.
\end{eqnarray*}

For later use we also present the following two-point functions,

\begin{eqnarray}
& &\langle 0 | \bar{\phi}'(\beta_1)\phi'(\beta_2) | 0 \rangle
= {\rm ln} w'(\beta_2-\beta_1),\nonumber\\
& &\langle 0 | \bar{\phi}'(\beta_1)\phi'(\beta_2) | 0 \rangle
= - {\rm ln} \bar{g}'(\beta_2-\beta_1),\nonumber
\end{eqnarray}

\noindent where

\begin{eqnarray}
& &w'(\beta) = \left[ g(\beta + \frac{ i \pi}{2})g(\beta - \frac{i \pi}{2})
\right] ^{-1}
= k^{-1} \frac{2\pi}{i(\beta-\frac{i\pi}{2})}, \nonumber\\
& &\bar{g}'(\beta ) = \left[ w'(\beta + \frac{ i \pi}{2})
w'(\beta - \frac{i \pi}{2}) \right] ^{-1}
= k^{2} \frac{\beta (\beta- i\pi) }{2\pi}.\nonumber
\end{eqnarray}

The locality of the operators $Z'_a(\beta)$ is best manifested by the
following specific operator products,

\begin{eqnarray*}
& &C_{ab}Z'_a(\beta+i\pi)Z'_b(\beta)=i,\\
& &Z'_a(\beta-i\pi)Z'_b(\beta)=iC_{ab},
\end{eqnarray*}

\noindent where $C_{ab}=\sigma_1$ (Pauli matrix) is the charge conjugation
matrix, and we shall also use its inverse $C^{ab}$: $C^{ac}C_{cb}=\delta^a_b$.

To end this review section, let us mention that the asymptotic operators
$Z_a(\beta)$ also satisfy an FZ algebra similar to (\ref{1}), but now with
the $S$-matrix replaced by

\begin{equation}
S_{ab}^{cd} (\beta) = - {S'}_{ab}^{cd}(-\beta), \label{S}
\end{equation}

\noindent and their asymptotic behavior is manifested by the following
operator product,

\begin{displaymath}
iZ_a(\beta_2)Z_b(\beta_1) = \frac{C_{ab}}{\beta_2-\beta_1-i\pi} + ...
\end{displaymath}.

Reasonably, these asymptotic operators should be regarded as appropriate
(weak) limits of the local operators $Z'_a(\beta)$ as $|t|
\rightarrow \infty$. However such limit must be taken in a nontrivial way
sence the transformation from the set of local operators to the set of
asymptotic operators is an improper unitary transformation.

\section{Boundary operators in $SU(2)$-invariant Thirring model}

In this and the forthcoming sections we shall concentrate on the
$SU(2)$-invariant Thirring model in the presence of reflecting boundaries.

In the presence of boundaries, one has the following
boundary reflection equation

\begin{equation}
Z'_a(\beta) |B \rangle = {R'}^{b}_{a}(\beta)Z'_b(-\beta) |B \rangle
\label{5}
\end{equation}

\noindent with the ``boundary state'' $|B\rangle$. This state should be
considered to be included in the Fock space ${\cal F}$, which is generated
by the ``boundary operator'' ${\rm exp}(\Psi_-)$ from the vacuum state
$| 0 \rangle$, {\rm i.e.}

\begin{equation}
| B \rangle = {\rm exp}(\Psi_-) | 0 \rangle,
\end{equation}

\noindent and the relation (\ref{5}) is an extension of the ZF algebra
(\ref{1}).

The consistence of (\ref{5}) and (\ref{1}) leads to the
boundary Yang-Baxter equation \cite{8}, \cite{10,10a}

\begin{eqnarray*}
& &{R'}^{c_2}_{a_2}(\beta_2){S'}^{c_1 d_2}_{a_1 c_2} (\beta_1 + \beta_2)
{R'}^{d_1}_{c_1}(\beta_1){S'}^{b_2 b_1}_{d_2 d_1} (\beta_1 - \beta_2)
\nonumber\\
& &\;\;\;\;\;\;=
{S'}^{c_1 c_2}_{a_1 a_2} (\beta_1 - \beta_2){R'}^{d_1}_{c_1}(\beta_1)
{S'}^{d_2 b_1}_{c_2 d_1} (\beta_1 + \beta_2){R'}^{b_2}_{d_2}(\beta_2)
\end{eqnarray*}

\noindent for the boundary reflection matrix ${R'}^{b}_{a}(\beta)$.
This equation, while supplied with the the boundary unitarity and boundary
crossing symmetry conditions \cite{8}

\begin{eqnarray*}
& &{R'}^{c}_{a}(\beta){R'}^{b}_{c}(-\beta)=\delta^b_a,\\
& &C^{ac}{R'}^{b}_{c}(-\frac{i \pi}{2}-\beta)= {S'}^{ab}_{cd}(2\beta)
C^{de}{R'}^{c}_{e}(-\frac{i \pi}{2}+\beta)
\end{eqnarray*}

\noindent give rise to the following solution,

\begin{eqnarray*}
{R'}^{b}_{a}(\beta)= \frac{R'(\beta)}{- \sqrt{4D_1 D_2+A^2} \beta - B}
\left(
\begin{array}{cc}
-A \beta - B & 2D_1 \beta \cr
2 D_2 \beta  & A \beta - B
\end{array}
\right),
\end{eqnarray*}
\begin{eqnarray*}
& &R'(\beta)=\frac{
\Gamma\left( \frac{ \frac{3i\pi}{2}-\beta}{2i\pi} \right)
\Gamma\left( \frac{ 2i\pi+\beta}{2i\pi} \right)
\Gamma\left( \frac{ \mu+\beta+i\pi}{2i\pi} \right)
\Gamma\left( \frac{ \mu-\beta+2i\pi}{2i\pi} \right)
}{
\Gamma\left( \frac{ \frac{3i\pi}{2}+\beta}{2i\pi} \right)
\Gamma\left( \frac{ 2i\pi-\beta}{2i\pi} \right)
\Gamma\left( \frac{ \mu-\beta+i\pi}{2i\pi} \right)
\Gamma\left( \frac{ \mu+\beta+2i\pi}{2i\pi} \right)
},\\
& &\mu=\frac{B}{\sqrt{4 D_1 D_2 + A^2}},
\end{eqnarray*}

\noindent where the constants $A,\;B,\;D_1,\;D_2$ are parameters related to
the boundary state $|B\rangle$.

In the rest of this article, we shall restrict ourselves to the relatively
simple case of $D_1=D_2=0$, which corresponds to the diagonal ${R'}$-matrix

\begin{eqnarray*}
{R'}^{b}_{a}(\beta)= R'(\beta)\left(
\begin{array}{cc}
1  & 0 \cr
0  & \frac{\mu +\beta}{\mu -\beta}
\end{array}
\right)
\end{eqnarray*}

\noindent with $\mu=B/A$.

Similar to $S'(\beta)$, $R'(\beta)$ can also be Riemann-Hilbert factorized,

\begin{eqnarray}
& &R'(\beta)
=\frac{f'(-\beta,\;\mu)}{f'(\beta,\;\mu)} \nonumber \\
& &f'(\beta,\;\mu)=\frac{
\Gamma\left( \frac{ 2i\pi-\beta}{2i\pi} \right)
\Gamma\left( \frac{ \mu-\beta+i\pi}{2i\pi} \right)
}{
\Gamma\left( \frac{ \frac{3i\pi}{2}-\beta}{2i\pi} \right)
\Gamma\left( \frac{ \mu-\beta+2i\pi}{2i\pi} \right)
}, \label{6}
\end{eqnarray}

\noindent where $f'(\beta,\;\mu)$ is analytic in the
lower half $\beta$-plane, {\it i.e.} $\Im m\beta < 0$.

For diagonal boundary reflection matrix, the boundary
reflection equation (\ref{5}) becomes (here no summation
is taken over $a$)

\begin{equation}
Z'_a(\beta) |B \rangle = {R'}^{a}_{a}(\beta)Z'_a(-\beta) |B \rangle,
\hskip 0.5truecm a=(+,\;-). \label{7}
\end{equation}

\noindent If $a=+$, we have from (\ref{6}-\ref{7}) that

\begin{eqnarray}
f'(\beta,\;\mu)Z'_+(\beta) | B \rangle =
(\beta \leftrightarrow -\beta). \label{8}
\end{eqnarray}

In order to study the property of the boundary operator $\Psi_-$, we first
introduce the following decomposition of the bosonic field $\phi'(\beta)$,

\begin{eqnarray*}
\phi'(\beta) = \phi'_+(\beta) + \phi'_0 + \phi'_-(\beta),
\end{eqnarray*}

\noindent where

\begin{eqnarray*}
& &\phi'_+(\beta) |0 \rangle = 0 = \langle 0 |\phi'_-(\beta),\\
& &\left[ \phi'_+(\beta_1),\;\phi'_-(\beta_2) \right] = - {\rm ln} g'(\beta_2
-\beta_1),\\
& &\left[ \phi'_0,\; \phi'_\pm(\beta) \right] = 0,\\
& &\left[ \phi'_0,\; \phi'_0 \right] = 0.
\end{eqnarray*}

\noindent These commutation relations imply the following two-point functions,

\begin{eqnarray*}
& &\langle 0 |\phi'_+(\beta_1)\phi'_-(\beta_2) |0\rangle =
- {\rm ln} g'(\beta_2 - \beta_1),\\
& &\langle 0| \phi'_0\phi'_\pm(\beta_2) | 0 \rangle = 0,\\
& &\langle 0| \phi'_0\phi'_0(\beta_2) | 0 \rangle = 0.
\end{eqnarray*}

\noindent We shall show in the next section that, while appropriately
regularized, the only nontrivial effect of $\phi'_0$ is to
decompose the Fock space into different sectors. The fact that $\phi'_0$
is independent of $\beta$ will also be clear in the next section.

Now let us make the ansatz

\begin{eqnarray}
& &\langle 0 | \Psi_- = 0,\nonumber\\
& &\left[ \Psi_-,\; \phi'_-(\beta) \right] = 0,\hskip 0.5truecm
\left[ \Psi_-,\; \phi'_0 \right] = 0,\nonumber\\
& &\left[ \Psi_-,\; \phi'_+(\beta) \right] = \kappa_- \phi'_-(-\beta)
+ \frac{1}{2} \gamma_-(-\beta),\nonumber\\
& &\left[ \gamma_-(\beta),\; {\rm everything} \right] = 0 \label{Psi-}.
\end{eqnarray}

\noindent for the operator $\Psi_-$.
Substituting the above ansatz into equation (\ref{8}), we get

\begin{eqnarray*}
\kappa_- = -1,\hskip 0.5truecm
{\rm exp}\left(- \frac{i}{2} \gamma_-(\beta) \right)
=g'(2\beta)^{1/2} f'(\beta,\;\mu).
\end{eqnarray*}

Let us check the consistency of the above ansatz and result with
the boundary reflection equation (\ref{7})
for $a=-$. This means that we should check the equality

\begin{equation}
(\mu - \beta) f'(\beta,\;\mu)
Z'_-(\beta) | B \rangle = ( \beta \leftrightarrow -\beta).
\label{13}
\end{equation}

\noindent From equation (\ref{Psi-}), we have

\begin{eqnarray}
& &{\rm e}^{i{\phi'}_+(\beta)} | B \rangle = f'(-\beta,\;\mu)
{\rm e}^{i{\phi'}_-(-\beta)} | B \rangle,\nonumber \\
& &{\rm e}^{-i {\bar{\phi}'}{}_{+} (\beta)} | B \rangle = -(2k)^{3/2}
\frac{(\mu + \beta + \frac{i\pi}{2})\beta}{4\pi^2} {\rm e}^{-i{\bar{\phi'}}
{}_{-}(-\beta)} | B \rangle. \label{14}
\end{eqnarray}

Using the explicit form of
$Z_-(\beta)$ and equation (\ref{14}), we find that equation (\ref{13}) is
equivalent to

\begin{equation}
\int_{C'}\frac{{\rm d}\;\gamma}{2\pi}
\frac{i\pi(\mu-\beta)(\mu+\gamma+ \frac{i\pi}{2}) \gamma}{
(\gamma-\beta-\frac{i\pi}{2})(\gamma-\beta+ \frac{i\pi}{2})(\gamma+\beta+
\frac{i\pi}{2})}\;
{\rm e}^{-i{\bar{\phi}'}{}_{-} (\gamma) - i{\bar{\phi}'}{}_{-} (-\gamma)}\;
| B \rangle = (\beta \leftrightarrow - \beta), \label{15}
\end{equation}

\noindent where the integration contour $C'$ encloses the poles $\gamma_1=
\beta-\frac{i\pi}{2}$ and $\gamma_2=-\beta-\frac{i\pi}{2}$ clockwise
but does not enclose the pole $\gamma_3=\beta+\frac{i\pi}{2}$. The last
equation can be directly verified by calculating the residues of the left
hand side at the poles $\gamma_1, \;\gamma_2$ and $\gamma_3$ respectively,
and then substracting from the sum of the first two residues the last one,
which results in an expression which is symmetric under $\beta
\leftrightarrow -\beta$.

Now let us mention that the same procedure as above can be
applied to obtained the left-handed boundary
operator ${\rm e}^{\Psi_+}$, which is used to construct the left-handed
boundary state $\langle B | = \langle 0 | {\rm e}^{\Psi_+}$, which
satisfies the boundary reflection equation

\begin{equation}
\langle B | {Z'}_a^{\ast} (\beta) = \langle B | {Z'}_a^{\ast} (-\beta)
{R'}_a^{\ast a}(-\beta), \label{16}
\end{equation}

\noindent where

\begin{eqnarray*}
& &{Z'}_a^{\ast} (\beta) = C_{ba}Z'_b(\beta + i\pi),\\
& &{R'}^{\ast b}_{a}(\beta)= {R'}^{b}_{a}(\beta)={R'}^{\ast}(\beta)
\left(
\begin{array}{cc}
\frac{\mu -\beta}{\mu + \beta} & 0 \cr
0  & 1
\end{array}
\right),\\
& &{R'}^{\ast} (\beta)=\frac{
\Gamma\left( \frac{ \frac{3i\pi}{2}-\beta}{2i\pi} \right)
\Gamma\left( \frac{ 2i\pi+\beta}{2i\pi} \right)
\Gamma\left( \frac{ \mu+\beta+i\pi}{2i\pi} \right)
\Gamma\left( \frac{ \mu-\beta}{2i\pi} \right)
}{
\Gamma\left( \frac{ \frac{3i\pi}{2}+\beta}{2i\pi} \right)
\Gamma\left( \frac{ 2i\pi-\beta}{2i\pi} \right)
\Gamma\left( \frac{ \mu-\beta+i\pi}{2i\pi} \right)
\Gamma\left( \frac{ \mu+\beta}{2i\pi} \right)
}.
\end{eqnarray*}

\noindent As before, ${R'}^{\ast} (\beta)$ can be factorized according to the
Riemann-Hilbert problem

\begin{eqnarray*}
& & {R'}^{\ast}(\beta) = \frac{ {f'}^{\ast} ( -\beta,\;\mu)}{
{f'}^{\ast} (\beta,\;\mu)},\\
& & {f'}^{\ast} (\beta,\;\mu) = \frac{
\Gamma\left( \frac{ 2i\pi-\beta}{2i\pi} \right)
\Gamma\left( \frac{ \mu-\beta+i\pi}{2i\pi} \right)}{
\Gamma\left( \frac{ \frac{3i\pi}{2}-\beta}{2i\pi} \right)
\Gamma\left( \frac{ \mu-\beta}{2i\pi} \right)}.
\end{eqnarray*}

As in the case of right-handed operators, we make the ansatz

\begin{eqnarray}
& & \Psi_+ | 0 \rangle = 0,\noindent\\
& &\left[ \Psi_+,\; \phi'_+(\beta) \right] = 0,\hskip 0.5truecm
\left[ \Psi_+,\; \phi'_0 \right] = 0,\noindent\\
& &\left[ \Psi_+,\; \phi'_-(\beta) \right] = \kappa_+ \phi'_+(-\beta
+2i\pi) + \frac{1}{2} \gamma_+(\beta-i\pi),\noindent\\
& &\left[ \gamma_+(\beta),\; {\rm everything} \right] = 0. \label{Psi+}
\end{eqnarray}

{}From the boundary reflection equation (\ref{16}) with $a=-$, we get

\begin{equation}
{f'}^{\ast} (-\beta,\;\mu) \langle 0 | {\rm e}^{\Psi_+} Z'_+(\beta + i\pi) =
( \beta \leftrightarrow - \beta)
\end{equation}

\noindent hence

\begin{equation}
\kappa_+ = 1,\hskip 0.5truecm
{\rm exp}\left(\frac{i}{2} \gamma_+(\beta) \right)= f'^{\ast}(\beta,\;\mu)
g'(2\beta)^{-1/2}.
\end{equation}

\noindent Furthermore, in analogy to equation (\ref{14}), we have

\begin{eqnarray*}
& &\langle B | {\rm e}^{ i {\phi'}{}_{-} (\beta +i\pi)} =
{f'}^{\ast} (\beta,\;\mu) \langle B | {\rm e}^{ i {\phi'}{}_{+}
( -\beta +i\pi)},\\
& &\langle B | {\rm e}^{-i {\bar{\phi}'}{}_{-} (\beta)} \sim
\frac{2i\pi-2\beta}{\mu-\beta+\frac{i\pi}{2}}
\langle B | {\rm e}^{ -i {\bar{\phi}'}{}_{+} ( -\beta + 2i\pi)}.
\end{eqnarray*}

\noindent where $\sim$ means ``equals up to normalization constant''.
The reflection equation in the case of $a=+$

\begin{eqnarray*}
\langle B | Z'_- (\beta + i\pi) = \langle B | Z'_- ( -\beta + i\pi)
\frac{\mu + \beta}{\mu - \beta} {R'}^{\ast} (- \beta)
\end{eqnarray*}

\noindent then is equivalent to

\begin{eqnarray*}
& &\int_{C^{\ast}} \frac{{\rm d}\;\gamma}{2\pi}
\frac{i\pi(\mu-\beta)(2 \gamma - 2 i\pi)}{
(\gamma+\beta-\frac{3i\pi}{2})(\gamma- \mu - \frac{i\pi}{2})
(\gamma-\beta - \frac{3i\pi}{2})(\gamma - \beta
- \frac{i\pi}{2})}\\
& &\hskip 1truecm
\times \langle B | {\rm e}^{-i{\bar{\phi}'}{}_{+}
(\gamma) - i{\bar{\phi}'}{}_{+} (-\gamma + 2i\pi)}
= (\beta \leftrightarrow - \beta),
\end{eqnarray*}

\noindent where the contour $C^{\ast}$ encloses only the pole $\gamma_1=
\beta+\frac{i\pi}{2}$ clockwise but does not enclose the poles
$\gamma_2=-\beta+\frac{3i\pi}{2}$, $\gamma_3=\mu+\frac{i\pi}{2}$ and
$\gamma_4=\beta+\frac{3i\pi}{2}$. This equality can be verified
using exactly the same technique as used to verify the equality (\ref{15}).

Before ending this section we would like to interpretate the functions
$\gamma_\pm(\beta)$ as appropriate vacuum expectation values under the
vacuum state $|0 \rangle$. From equations (\ref{Psi-}) and
(\ref{Psi+}) we can easily see that

\begin{eqnarray*}
& &\langle 0 | \phi'_+(\beta) pPsi_- | 0 \rangle = \frac{1}{2}
\gamma_-(-\beta),\\
& &\langle 0 | \Psi_+ \phi'_-(\beta)  | 0 \rangle = \frac{1}{2}
\gamma_+(\beta-i\pi).
\end{eqnarray*}

\noindent On the other hand, defining $F=\left[ \Psi_+,\;\Psi_- \right]$,
we can easily
calculate the following relations using the ansatzs (\ref{Psi-}) and
(\ref{Psi+}),

\begin{eqnarray}
& &\left[ F,\;\phi'_+(\beta) \right] = - \phi'_+ (\beta+2i\pi) - \frac{1}{2}
\gamma_+(-\beta-i\pi),\nonumber\\
& &\left[ F,\;\phi'_-(\beta) \right] = \phi'_- (\beta-2i\pi) - \frac{1}{2}
\gamma_-(\beta-2i\pi). \label{shift}
\end{eqnarray}

\noindent Thus the functions $\gamma_\pm(\beta)$ are also equivalent to the
following vacuum expectation values

\begin{eqnarray*}
& &\langle 0 | \phi'_+(\beta) F | 0 \rangle = \frac{1}{2}
\gamma_+(-\beta-i\pi),\\
& &\langle 0 | F\phi'_-(\beta)  | 0 \rangle = -\frac{1}{2}
\gamma_-(\beta-2i\pi).
\end{eqnarray*}

Notice the peculiar effect of $F$ on $\phi'_\pm$ (see
equation (\ref{shift})). Besides changing the signs of $\phi'_\pm(\beta)$
and shifting by scalar functions, $F$ also shift the rapidities of
$\phi'_\pm(\beta)$ by $\\pm 2i\pi$. It is possible that behind this
phenomenon there lies some deep reasoning. However at this moment we
cannot say anything about that.

\section{Boundary operators versus $q$-oscillators}

The bosonic field $\phi'(\beta)$ essentially carries singularities.
Therefore, it cannot be expanded into an infinite sum of oscillator modes
as the usual free field does. However, we can perform an oscillator
realization by the ultraviolate regularization introduced by Lukuyanov
\cite{6}. To this end the first thing to do is to introduce the
regularization parameter $\epsilon$ such that

\begin{displaymath}
-\frac{\pi}{\epsilon} \leq | \beta | \leq \frac{\pi}{\epsilon}
\end{displaymath}

\noindent and consider the ``regularized field'' $\phi'_\epsilon$.
Then the boson $\phi'$ can be viewed as proper limit of $\phi'_\epsilon$
as $\epsilon \rightarrow 0$.

In analogy to (\ref{phiphi}), we have the following commutation relation,

\begin{eqnarray*}
\left[\phi'_\epsilon(\beta_1),\;\phi'_\epsilon(\beta_2)\right]={\rm ln}\;
S'_\epsilon(\beta_2-\beta_1),
\end{eqnarray*}

\noindent where

\begin{eqnarray}
& &S'_\epsilon(\beta)={\rm exp} \left(\frac{- i \epsilon \beta}{2}\right)
\frac{g'_\epsilon(-\beta)}{g'_\epsilon(\beta)},\nonumber\\
& &g'_\epsilon(\beta)=(1-q)^{1/2}\;\frac{\Gamma_q
\left(\frac{2i\pi-\beta}{2i\pi}\right)}{\Gamma_q
\left(\frac{i\pi-\beta}{2i\pi}\right)},\hskip 0.5truecm \left(q={\rm exp}
(-2\pi\epsilon)\right)\nonumber\\
& &\Gamma_q(x)=(1-q)^{1-x} \prod^{\infty}_{k=1}
\frac{1-q^k}{1-q^{x+k-1}}.\nonumber
\end{eqnarray}

The oscillator expansion of the regularized boson $\phi'_\epsilon(\beta)$
reads

\begin{eqnarray*}
\phi'_\epsilon(\beta)=\phi'_{\epsilon 0}(\beta)
+ {\phi'_\epsilon}_+ (\beta) + {\phi'_\epsilon}_- (\beta),
\end{eqnarray*}

\noindent where

\begin{eqnarray}
& &\phi'_{\epsilon 0}(\beta) =
-\frac{1}{\sqrt{2}}(Q-\epsilon \beta P),\nonumber\\
& &i {\phi'_\epsilon}_+ (\beta) = -\sum^{\infty}_{m=1} \frac{a'_m}{{\rm sh}
(\pi m \epsilon)} {\rm exp}(i \pi \epsilon \beta),\nonumber\\
& &i {\phi'_\epsilon}_- (\beta) = \sum^{\infty}_{m=1} \frac{a'_{-m}}{{\rm sh}
(\pi m \epsilon)} {\rm exp}(- i \pi \epsilon \beta),\nonumber\\
& &\left[P,\;Q\right]=-i,\nonumber\\
& &\left[a'_m,\;a'_n\right]=\frac{{\rm sh}\frac{\pi m \epsilon}{2}
{\rm sh}{\pi m \epsilon}}{m}\;{\rm exp}\left(-\frac{\pi |m| \epsilon}{2}
\right) \delta_{m+n,\;0}.\label{osci}
\end{eqnarray}

\noindent Here the Fock space ${\cal F}_\epsilon$ is defined via the ``vacuum
state'' $|0 \rangle_p$ on which the oscillators act as follows \cite{5},

\begin{eqnarray*}
& &a'_m |0 \rangle_p = 0, \hskip 0.5truecm (p>0) \\
& &P |0 \rangle_p = p |0 \rangle_p.
\end{eqnarray*}

\noindent Notice that the action of ${\rm e}^Q$ on the state $|0 \rangle_p$
shifts the eigenvalue of $P$ by 1 and does not affect the actions of
$a'_\pm m$. So the space ${\cal F}_\epsilon$ has the structure

\begin{eqnarray*}
{\cal F}_\epsilon = \oplus_p {\cal F}_p,
\end{eqnarray*}

\noindent where the ``vacuum state'' of ${\cal F}_p$ is $|0\rangle_p$. In
the limit of $\epsilon \rightarrow 0$, the space ${\cal F}_\epsilon$ tends
to ${\cal F}$, with the vacuum $|0\rangle_0 \rightarrow |0 \rangle$.

The generators of ZF algebra in the presence of ultraviolate
regularization read \cite{2a,6}

\begin{eqnarray}
& &Z'_{\epsilon +}(\beta)
= {\rm e}^{\frac{i \epsilon \beta}{4}}\;V'_{\epsilon}
(\beta) = {\rm e}^{\frac{i \epsilon \beta}{4}}\;{\rm e}^{i \phi'_{\epsilon}
(\beta)},\nonumber\\
& &Z'_{\epsilon -}(\beta)
= i {\rm e}^{-\frac{i \epsilon \beta}{4}}\;
\left({\rm e}^{-\frac{\pi \epsilon}{4}} \chi'_\epsilon
V'_{\epsilon}(\beta) + {\rm e}^{\frac{\pi \epsilon}{4}}
V'_{\epsilon}(\beta) \chi'_\epsilon \right), \label{ze}
\end{eqnarray}

\noindent which satisfy ZF algebra with the ``regularized'' $S$-matrix

\begin{eqnarray*}
& &{S'_\epsilon}^{++}_{++}(\beta)={S'_\epsilon}^{--}_{--}(\beta)
={S'}_\epsilon (\beta),\\
& &{S'_\epsilon}^{+-}_{+-}(\beta)={S'_\epsilon}^{-+}_{-+}(\beta)
=-{S'}_\epsilon (\beta)
\frac{{\rm sh} \frac{i \epsilon \beta}{2}}{{\rm sh} \frac{i \epsilon
(i \pi + \beta)}{2}},\\
& &{S'_\epsilon}^{+-}_{-+}(\beta)={S'_\epsilon}^{-+}_{+-}(\beta)
=-{S'}_\epsilon (\beta)
\frac{{\rm sh} \frac{\pi \epsilon}{2}}{{\rm sh} \frac{i \epsilon
(i \pi + \beta)}{2}}.
\end{eqnarray*}

\noindent Correspondingly, the diagonal ``ultraviolate regularized''
$R'$-matrix is

\begin{eqnarray*}
{R'_\epsilon}^{b}_{a}(\beta)= R'_\epsilon(\beta)\left(
\begin{array}{cc}
1  & 0 \cr
0  & \frac{ {\rm sh} \frac{i \epsilon (\mu +\beta)}{2}}{ {\rm sh}
\frac{i \epsilon (\mu -\beta)}{2}}
\end{array}
\right),
\end{eqnarray*}

\noindent where

\begin{eqnarray*}
R'_\epsilon(\beta)= \frac{
\Gamma_{q^2}\left( \frac{ \frac{3i\pi}{2}-\beta}{2i\pi} \right)
\Gamma_{q^2}\left( \frac{ 2i\pi+\beta}{2i\pi} \right)
\Gamma_{q}  \left( \frac{ \mu+\beta+i\pi}{2i\pi} \right)
\Gamma_{q}  \left( \frac{ \mu-\beta+2i\pi}{2i\pi} \right)
}{
\Gamma_{q^2}\left( \frac{ \frac{3i\pi}{2}+\beta}{2i\pi} \right)
\Gamma_{q^2}\left( \frac{ 2i\pi-\beta}{2i\pi} \right)
\Gamma_{q}  \left( \frac{ \mu-\beta+i\pi}{2i\pi} \right)
\Gamma_{q}  \left( \frac{ \mu+\beta+2i\pi}{2i\pi} \right)
}.
\end{eqnarray*}

\noindent In analogy to equation (\ref{6}), we have the following
Riemann-Hilbert problem,

\begin{eqnarray*}
& &R'_\epsilon (\beta)
=\frac{f'_\epsilon(-\beta,\;\mu)}{f'_\epsilon(\beta,\;\mu)} \\
& &f'_\epsilon(\beta,\;\mu)=\frac{
\Gamma_{q^2}\left( \frac{ 2i\pi-\beta}{2i\pi} \right)
\Gamma_{q}  \left( \frac{ \mu-\beta+i\pi}{2i\pi} \right)
}{
\Gamma_{q^2}\left( \frac{ \frac{3i\pi}{2}-\beta}{2i\pi} \right)
\Gamma_{q}  \left( \frac{ \mu-\beta+2i\pi}{2i\pi} \right)
},
\end{eqnarray*}

\noindent where $f'_\epsilon(\beta,\;\mu)$ is analytic in the
lower half $\beta$-plane.

The ansatz for $\Psi_-$ now reads

\begin{equation}
\Psi_-=\sum_{n=1}^{\infty}\frac{ n{\rm exp} \left( \frac{\pi
n \epsilon}{2} \right) \alpha_n}{2 {\rm sh} \frac{\pi n \epsilon}{2}
{\rm sh} \pi n \epsilon}\;{a'}_{-n}^2
+ \sum_{n=1}^{\infty} \frac{n {\rm exp} \left(\frac{\pi n \epsilon}{2}\right)
}{{\rm sh} \frac{\pi n \epsilon}{2}}
\lambda_n {a'}_{-n},  \label{10}
\end{equation}

\noindent where $\alpha_{n},\;\lambda_{n}$ are
parameters to be determined.
This form of ansatz for $\Psi_-$ was first introduced by the
paper \cite{9} in which similar problem for $XXZ$-spin chain was considered.
In the same paper the rational limit of $XXX$-spin chain was also taken.
Here we adopt this ansatz to achieve an oscillator realization of boundary
operators in a field theoretic model.

Following the fundamental commutation
relation (\ref{osci}) of the oscillators we have

\begin{eqnarray}
& &{\rm e}^{-\Psi_-}a'_{n}{\rm e}^{\Psi_-}=a'_n + \alpha_n a'_{-n} +
\lambda_n {\rm sh} \pi n \epsilon,\nonumber \\
& &{\rm e}^{-\Psi_-}a'_{-n}{\rm e}^{\Psi_-}=a'_{-n},\hskip 0.5truecm
n>0. \label{11}
\end{eqnarray}

\noindent Substituting equations (\ref{ze}), (\ref{10}-\ref{11}) into
(\ref{8}), we get

\begin{eqnarray}
& &\alpha_n=-1,\nonumber \\
& &\lambda_n=-\frac{q^{\frac{\mu + i\pi}{2i\pi}n}}{n\left(1+q^{\frac{n}{2}}
\right)} + \theta_n \frac{q^{\frac{n}{2}}-q^{\frac{n}{4}}}{n(1+q^{\frac{n}{2}}
)},\label{12}
\end{eqnarray}

\noindent where

\begin{displaymath}
\theta_n=\left\{
\begin{array}{cc}
0,&\;\;\;m\;{\rm odd} \cr
1,&\;\;\;m\;{\rm even}
\end{array}
\right.
\end{displaymath}

The regularized form of equation (\ref{13}) reads

\begin{equation}
{\rm sh} \frac{i \epsilon (\mu - \beta)}{2} f'_\epsilon(\beta,\;\mu)
Z'_{\epsilon -} (\beta) | B \rangle = ( \beta \leftrightarrow -\beta).
\label{133}
\end{equation}

\noindent Using exactly the same procedures as in the non-regularized case
we can show that the solution (\ref{10},\ref{12})
for the boundary operator $\Psi_-$
is consistent with equation (\ref{133}). Similarly, making the ansatz

\begin{eqnarray*}
\Psi_+=\sum_{n=1}^{\infty} \frac{ n {\rm exp} \left( \frac{\pi
n \epsilon}{2} \right) \sigma_n}{2 {\rm sh} \frac{\pi n \epsilon}{2}
{\rm sh} \pi n \epsilon}\;{a'}_{n}^2
+ \sum_{n=1}^{\infty} \frac{n {\rm exp} \left(\frac{\pi n \epsilon}{2}\right)
}{{\rm sh} \frac{\pi n \epsilon}{2}}
\rho_n {a'}_{n}
\end{eqnarray*}

\noindent for $\Psi_+$, we find the following consistent coefficients,

\begin{eqnarray*}
& &\sigma_n=-q^n,\\
& &\rho_n=-\frac{q^{\frac{\mu + i\pi}{2i\pi}n}}{n \left(1+q^{\frac{n}{2}}
\right)} + \theta_n \frac{q^{\frac{3n}{4}}-q^{n}}{n(1+q^{\frac{n}{2}})
}.
\end{eqnarray*}

\section{Reflection equations for asymptotic operators}

Having finished the construction of boundary operators using the reflection
equations of local ZF generators, let us now turn to the other kind of ZF
generators, {\it i.e.} the asymptotic generators.
As mentioned in Section 2, these operators also satisfy ZF algebra (\ref{1}),
with a different $S$-matrix given by equation (\ref{S}). Moreover,
they can also be bosonized using a boson $\phi$
which is closely related to the free boson $\phi'$ used in the last section.
Therefore the operators $Z_a(\beta)$ can also effectively act on the
Fock space of the free boson $\phi'$, and hence there arise the necessity
of checking the consistency between the boundary operators obtained above
and the boundary reflection equations for the operators $Z_a(\beta)$.

Let us proceed in some more detail. First, using the boundary Yang-Baxter
equation, boundary unitarity and the boundary crossing symmetry

\begin{displaymath}
C^{ac}R^{b}_{c}(-\beta + \frac{i\pi}{2})
={S}^{ab}_{cd}(2\beta)
C^{de}R^{c}_{e}(\beta + \frac{i\pi}{2})
\end{displaymath}

\noindent corresponding
to the asymptotic operators, we obtain the (diagonal) boundary reflection
matrix

\begin{eqnarray*}
& &{R}^{b}_{a}(\beta)= R(\beta)\left(
\begin{array}{cc}
1  & 0 \cr
0  & \frac{\nu + \beta}{\nu - \beta}
\end{array}
\right),\\
& &R(\beta)=\frac{
\Gamma\left( \frac{-\beta}{2i\pi} \right)
\Gamma\left( \frac{ \frac{i\pi}{2}+\beta}{2i\pi} \right)
\Gamma\left( \frac{ \nu-\beta+i\pi}{2i\pi} \right)
\Gamma\left( \frac{ \nu+\beta+2i\pi}{2i\pi} \right)
}{
\Gamma\left( \frac{ \beta}{2i\pi} \right)
\Gamma\left( \frac{ \frac{i\pi}{2}-\beta}{2i\pi} \right)
\Gamma\left( \frac{ \nu+\beta+i\pi}{2i\pi} \right)
\Gamma\left( \frac{ \nu-\beta+2i\pi}{2i\pi} \right)
}\\
& &\;\;\;\;\;=\frac{
f(-\beta,\;\nu)}{f(\beta,\;\nu)},\\
& &f(\beta,\;\nu)=\frac{
\Gamma\left( \frac{ \frac{i\pi}{2}-\beta}{2i\pi} \right)
\Gamma\left( \frac{ \nu-\beta+2i\pi}{2i\pi} \right)
}{
\Gamma\left( \frac{-\beta}{2i\pi} \right)
\Gamma\left( \frac{ \nu-\beta+i\pi}{2i\pi} \right)
}.
\end{eqnarray*}

In order that the operators $Z_a(\beta)$ and $Z'_a(\beta)$ have the same
boundary states we require that the boundary parameters $\mu$ and $\nu$
be related as

\begin{displaymath}
\nu=\mu-\frac{i\pi}{2}.
\end{displaymath}

\noindent Lukuyanov also gave the bosonization formula for the operators
$Z_a(\beta)$, which read

\begin{eqnarray*}
& &Z_+(\beta)=V(\beta) \equiv {\rm e}^{ i\phi(\beta)},\\
& &Z_-(\beta)=i (\chi V(\beta) + V(\beta)\chi),
\end{eqnarray*}

\noindent where $\chi$ is defined through the matrix elements

\begin{eqnarray*}
& &\langle u | \chi | v \rangle = \xi \langle u |
\int_C \frac{{\rm d}\gamma}{2\pi}
\bar{V}(\gamma) | v \rangle,\\
& &\bar{V}(\gamma) = :{\rm e}^{i \bar{\phi} (\gamma)}: =
:{\rm e}^{i \bar{\phi} (\gamma + \frac{i\pi}{2}) +
i \bar{\phi} (\gamma - \frac{i\pi}{2})}:,
\end{eqnarray*}

\noindent and the principles for choosing the integration contour $C$
are the same for choosing the contour $C'$, and the boson $\phi$ is
connected to $\phi'$ via

\begin{eqnarray}
& &\phi_+(\beta) = - {\phi'}_+ (\beta-\frac{i\pi}{2}), \nonumber\\
& &\phi_-(\beta) = - {\phi'}_- (\beta+\frac{i\pi}{2}). \label{17}
\end{eqnarray}

\noindent Using equations (\ref{14}) and (\ref{17}) we can get

\begin{eqnarray*}
& &{\rm e}^{i\phi_+(\beta)} |B \rangle = \frac{g'(-2\beta + i\pi)}{f'(-\beta+
\frac{i\pi}{2})}\; {\rm e}^{i\phi_- (-\beta)} | B \rangle,\\
& &{\rm e}^{- i \bar{\phi}_+ (\beta)} |B \rangle
\sim \frac{2\beta}{\mu+\beta}\; {\rm e}^{- i\bar{\phi}_- (-\beta)} | B \rangle.
\end{eqnarray*}

\noindent Using the method of Sections 3 and 4 we can verify that the
boundary state $|B\rangle$ again solve the boundary reflection equation

\begin{eqnarray*}
Z_a(\beta) |B\rangle = R^{b}_{a}(\beta) Z_b(-\beta) |B\rangle
\end{eqnarray*}

\noindent for asymptotic operators with diagonal reflection matrix $R$.
Simllarly we can check that the boundary state obtained from the operator
${\rm e}^{\Psi_+}$ satisfy the left-handed reflection equation

\begin{eqnarray*}
\langle B|Z_a(\beta + i\pi) = \langle B | Z_b (-\beta +i\pi)
R^{\bar{b}}_{\bar{a}}(-\beta).
\end{eqnarray*}

\noindent This finishes the full proof for the consistency between the
boundary states and the boundary reflection equations for the asymptotic
operators.

\section{Conclusion}

In this article we derived the explicit expression for the boundary
operators (states) using the technique of bosonization via deformed bosonic
fields and oscillators. This is just a partial result for a more involved but
self-contained project of calculating the multi-point correlation functions
and form factors in the boundary scattering theory. We hope to present the
full result elsewhere. We also hope to consider some different models
such as Sine-Gordon model, {\it etc}. The more general problems such as
the connections between deformed bosonic fields and deformed Virasoro algebra
\cite{aa}, the boundary Knizhnik-Zamolodchikov equation and form factor
axioms \cite{9a} are all worth of further study.

\bigskip
\medskip

\noindent{\bf \Large Acknowledgement}

\bigskip

The authors thank O.Foda for transfer the manuscript \cite{9} to us.


\newpage

\begin{thebibliography}{15}

\bibitem{1} M Jimbo, T Miwa, {\it Algebraic Analysis of Solvable Lattice
Models}, RIMS Preprint-981 (1994)

\bibitem{2}E Date, M Okado, Int J Mod Phys A9 (1994) 399

\bibitem{2a}O Foda, M Jimbo, T Miwa, K Miki and A Nakayashiki, J Math Phys 35
(1994) 13

\bibitem{2b}M Idzumi, K Iohara, M Jimbo, T Miwa, T Nakashima and T Tokihiro,
Int J Mod
Phys A8 (1993) 1479

\bibitem{2c}M Jimbo, T Miwa and Y Ohta, Int J Mod Phys A8 (1993) 1457

\bibitem{2d}Bougurzi, Weston, {\it N-point correlation functions of the
spin-1 XXZ model}, preprint CRM-1896, Univ de Montreal (1993)

\bibitem{4} V G Drinfel'd, {\it Quantum Group}, Proc ICM-86 (Berkeley, CA), 1
(1987) 798

\bibitem{5} I B Frenkel, N H Jing, Proc Nat'l Acad Sci USA 85 (1988) 9373

\bibitem{4a} A Kato, Y H Quano and J Shirashi, Commun Math Phys 157 (1993) 119

\bibitem{4b} A Matsuo, Commum Math Phys 160 (1994) 33

\bibitem{4c} J Shirashi, Phys Lett A171 (1992) 243

\bibitem{6} S Lukuyanov, {\it Free field representation for massive integrable
models}, RU-93-30, hep-th/9307196; Phys Lett B325 (1994) 409

\bibitem{a} S Lukuyanov, S L Shatashvilli, Phys Lett B298 (1993) 111

\bibitem{aa} S Lukuyanov, Y. Pugal, {\it Bosonization of ZF algebras:
direction toward deformed Virasoro algebra}, RU-94-41, hep-th/9412128

\bibitem{7} F A Smirnov, {\it Form factors in complete integrable models of
quantum field theory}, World Scientific (Singapore) 1992

\bibitem{7a} F A Smirnov, {\it Lectures on integrable massive
models of quantum field
theory}, in ``Introduction to quantum group and integrable massive models of
quantum field theory'', Nankai Lectures on Mathematical Physics, edited by
M L Ge and B H Zhao, World Scientific (1990)

\bibitem{8} S Ghoshal, A Zamolodchikov, Int J Mod Phys A9 (1994) 3841

\bibitem{9} M Jimbo, R Kedem, T Kojima, T kojima, H Koono and T Miwa,
{\it XXZ-chain with a boundary},
hep-th/9411112 (to be published in Nucl. Phys. B)

\bibitem{9a} M Jimbo, R Kedom, H Konno, T Miwa and R Weston,
{\it Difference equations in spin chain with a boundary}, RIMS-1005, CRM-2246,
hep-th/9502060

\bibitem{10} I V Cherednik, Theore Math Phys 61 (1984) 977

\bibitem{10a} E K Sklyanin, J Phys A: Math Gen 21 (1988) 2375

\end{thebibliography}
\end{document}